\documentclass[12pt]{article}
\usepackage[dvips]{graphicx}
\title{Liouville Operator Approach to Symplecticity-Preserving Renormalization
 Group Method}
\author{Shin-itiro Goto$^a$\footnote{e-mail:~sgoto@amp.i.kyoto-u.ac.jp},
\quad Kazuhiro Nozaki$^b$ \\
{\small $a$ Department of Applied Mathemaics and Physics,}\\ 
{\small Kyoto University, Kyoto, 606-8501, Japan}\\
{\small $b$ Department of Physics, }\\
{\small Nagoya University, Nagoya 464-8602, Japan}
}
\date{}
\makeindex

\begin{document}
\maketitle

\newcommand{\beq}{\begin{equation}}
\newcommand{\beqa}{\begin{eqnarray}}
\newcommand{\eeq}{\end{equation}}
\newcommand{\eeqa}{\end{eqnarray}}
\newcommand{\non}{\nonumber}
\newcommand{\lb}{\label}
\newcommand{\fr}[1]{(\ref{#1})}
\newcommand{\cc}{\mbox{c.c.}}
\newcommand{\nr}{\mbox{n.r.}}
\newcommand{\tx}{\widetilde{x}}
\newcommand{\tg}{\widetilde{g}}
\newcommand{\hx}{\widehat{x}}
\newcommand{\tA}{\widetilde A}
\newcommand{\tB}{\widetilde B}
\newcommand{\tK}{\widetilde K}
\newcommand{\tc}{\widetilde c}
\newcommand{\tAc}{{\widetilde A}^{*}}
\newcommand{\tphi}{{\widetilde \phi}}
\newcommand{\btA}{\mbox{\boldmath {$\widetilde A$}}}
\newcommand{\bA}{\mbox{\boldmath {$A$}}}
\newcommand{\bC}{\mbox{\boldmath {$C$}}}
\newcommand{\bu}{\mbox{\boldmath {$u$}}}
\newcommand{\bN}{\mbox{\boldmath {$N$}}}
\newcommand{\bZ}{\mbox{\boldmath {$Z$}}}
\newcommand{\bR}{\mbox{\boldmath {$R$}}}
\newcommand{\ve}{{\varepsilon}}
\newcommand{\e}{\mbox{e}}
\newcommand{\aCos}{\mbox{Cos}^{-1}}
\newcommand{\aTan}{\mbox{Tan}^{-1}}
\newcommand{\mapright}[1]{%
  \smash{\mathop{%
     \hbox to 2cm{\rightarrowfill}}\limits^{#1}}}
\newcommand{\mapleft}[1]{%
  \smash{\mathop{%
     \hbox to 2cm{\leftarrowfill}}\limits^{#1}}}
\def\lmapdown#1{\Big\downarrow
  \llap{$\vcenter{\hbox{$\scriptstyle#1\,$}}$ }}
\def\rmapdown#1{\Big\downarrow
  \rlap{$\vcenter{\hbox{$\scriptstyle#1\,$}}$ }}

%%%%%%%%%%%%%%%%
\begin{abstract}
%%%%%%%%%%%%%%%%%
We present a method to construct symplecticity-preserving 
renormalization group maps by using the Liouville operator,
and obtain correctly reduced symplectic maps describing their
long-time behavior even when a resonant island chain appears.
%It is found that the modulational instability in
%   the reduced map triggers random wandering of orbits around
%   some homoclinic-like manifolds, which is understood as
%the Bernoulli shifts.

PACS-1995:
03.20.+i ,
%(classical mechanics of discrete systems: general mathematical aspects)
47.20.Ky,
%(Nonlinearity (including bifurcation theory)
02.30.Mv,
%(Approximations and expansions)
64.60.Ak.
%(Renormalization-group, fractal, and percolation studies of phase transitions)

\end{abstract}
%%%%%%%%%%%%%%

%%%%%%%%%%%
\pagebreak
%%%%%%%%%%%

%%%%%%%%%%%%%%%%%%%%%%
\section{Introduction}
%%%%%%%%%%%%%%%%%%%%%%
There has been a long history to study an asymptotic solution of
Hamiltonian flows by means of singular perturbation methods such as
the averaging method and the method of multiple time-scales.
A Hamiltonian flow can be reduced to a symplectic discrete map
called the Poincar\'e map,
which has the lower dimension than the original flow and is, therefore,
extensively studied \cite{LL91}\cite{AA}. 
However, neither the averaging method
nor the method of  multiple time-scales may  be immediately applicable to
symplectic maps.

The perturbative renormalization group (RG) method developed recently
may be a useful tool to tackle asymptotic
behaviours of discrete maps as well as flows.
The original RG method is an asymptotic singular perturbation technique
developed for diffential equations \cite{CGO96}. Secular or divergent terms
of  perturbation solutions of differential equations are
removed by renormalizing integral constants of the lowest order solution.
The RG method is reformulated on the basis of a naive renormalization
transformation and the Lie group \cite{GMN99}.
This reformulated RG method based on the Lie group is easy to apply to
discrete systems, by which asymptotic expansions of unstable
manifolds of some chaotic discrete systems are obtained \cite{GN01Prog}.
The extension of the RG method to discrete symplectic systems is
not trivial because the symplectic struture is easily broken
in naive RG equations (maps)
as shown in Ref.\cite{GN01JPSJ}, while the application of the RG method to
Hamiltonian flows does not cause such a problem as the broken
symplectic symmetry except a special case \cite{YN98}.

The application of the RG method to some non-symplectic discrete
systems has been attempted in the framework of the envelope method \cite{KM98}.
Howevever, the method, if applied to a symplectic map, would give only
a naive RG map which breaks the symplectic symmetry.
To recover the symmetry breaking,  
not only the paper \cite{GN01JPSJ} has introduced a process of 
the ``exponentiation''
which gives us analytical expressions of some physical quantities 
\cite{GN01JPSJ}\cite{Tze01}, 
but also another paper \cite{GNY02} has used a 
symplectic integrator by taking the time-continuous limit of the time-step 
parameter. Furthermore, S.I. Tzenov et al have showed that the exponentiation procedure can be successfully applied in the H\'enon map\cite{TD03}.
However this procedure may seem somewhat artifical, and resonant islands have
never been studied by the RG method.

The main purpose of the present paper is to present
a general RG procedure to preserve the symplectic structure in
RG maps and to obtain correctly reduced symplectic RG maps. 
In this paper, this procedure
is called a symplecticity-preserving RG procedure, 
which consists of the following two steps.
First, using the reformulated RG method \cite{GMN99}, we get a naive
RG map near elliptic of some symplectic discrete systems.  
The naive RG map preserves the
symplectic symmetry only approximately and fails to describe
a long-time behavior of the original maps.
Second, in order to recover the symplectic symmetry by means of taking account
of higher orders in small parameters in the naive perturbation result. Then
we equate the naive RG map to the appropriately discretized Hamiltonian flow,
which is done by using the Liouville operator. 
On this procedure we identify the small parameter $\ve$ with the time-step, 
which yields a canonical equation in the limit of $\ve\to 0$. 
This process and a symplectic RG map are
called, respectively, a symplecticity-preserving procedure and a 
symplecticity-preserving RG map.
A reduction of the Liouville equation in 
a non-integrable Hamiltonian flow has been developed in Ref.\cite{Tze02}.

In section 2, a long-time behaviour
of a simple linear map is analyzed  in order to elucidate both the
broken symplectic symmetry in a naive RG map and our regularization
process.
In section 3 , a symplecticity-preserving RG map is obtained near
elliptic fixed points of a two-dimensional nonlinear
symplectic map even when a resonant island chain appears.
In section 4, we mention the advantages of our newly developed RG method as 
the summary of this paper. 

%%%%%%%%%%%%%%%%%%%%%%%%%%%%%%%%%%%%%%%%%%%%%%%%%%%%%%
\section{Linear symplectic map}
%%%%%%%%%%%%%%%%%%%%%%%%%%%%%%%%%%%%%%%%%%%%%%%%%%%%%%%
It may be instructive to analyze a linear symplectic map, which
is exactly solvable, $(x^n,y^n)\mapsto (x^{n+1},y^{n+1})$:
\beqa
x^{n+1}&=&x^n+y^{n+1}\non\\
y^{n+1}&=&y^n-ax^n+2\ve Jx^n.\non
\eeqa
This map can be rewritten as
\beq
L_\theta x^n=\ve 2 J x^n,
\lb{eqn:linear}
\eeq
where $\ve$ is the small parameter, $J$ is the real parameter, $\theta$ and $L_\theta x^n$ are defined as 
\beqa
\cos\theta&\equiv& (1-a/2),
\lb{def:theta}\\
L_\theta x^n&\equiv&x^{n+1}-2 x^n \cos\theta +x^{n-1},
\lb{def:L}
\eeqa
respectively.
We assume that the map \fr{eqn:linear} has an elliptic fixed point at the
origin $(0,0)$.
%where $\cos\theta=1-J/2$ 
%and $0<J<2$ is assumed.
The linear map \fr{eqn:linear} has the following exact solution $x_E^n$ :
\beqa
x_E^n&=&A\exp\bigg(i\arccos(\cos\theta+\ve J)n\bigg)+\cc,\non\\
     &=&A\exp\Bigg[i\Bigg(\theta +\ve\frac{-J }{\sin\theta}
                +\ve^2\frac{-\cos\theta}{2\sin\theta}
                          \bigg(\frac{J}{\sin\theta}\bigg)^2
                          +\cdots \Bigg)n\Bigg]+\cc,
\lb{eqn:linear_exact_x_sol}
\eeqa
where $A(\in\bC)$ is the complex ``integration'' constant
and $\cc$ stands for the complex conjugate to the preceding terms.

Let us derive an asymptotic solution of the map \fr{eqn:linear}
for small $\ve$ by means of the RG method.
Substituting the expansion
\beq
x^n=x^{(0)n}+\ve x^{(1)n}+\ve^2x^{(2)n}+\cdots, \non
\eeq
into Eq.\fr{eqn:linear}, we have
$$
L_\theta x^{(0)n}=0,\quad L_\theta x^{(1)n}=2Jx^{(0)n},\quad L_\theta x^{(2)n}=2Jx_n^{(1)n}
\cdots.
$$
and
\beqa
x^{(0)n}&=&A\exp(i\theta n)+\cc,\non\\
x^{(1)n}&=&\frac{-iJA}{\sin\theta}n\exp(i\theta n)+\cc,\non\\
x^{(2)n}&=&\frac{-J^2A}{2\sin^2\theta}\bigg(
     n^2+i\frac{\cos\theta}{\sin\theta}n\bigg)
\exp(i\theta n)+\cc,\non
\eeqa
where $A(\in\bC)$ is the integration constant.
To remove secular terms ($\propto n, n^2$),
we introduce the renormalization transformation
$A\mapsto A^n$ \cite{GMN99}:
\beq
A^n\equiv A+\ve\frac{-iJA}{\sin\theta}n
             +\ve^2\frac{-J^2A}{2\sin^2\theta}\bigg(
                 n^2+i\frac{\cos\theta}{\sin\theta}n\bigg)
             +{\cal O}(\ve^3),
\lb{eqn:linear_RGT}
\eeq
A discrete version of the RG equation is just the first order
difference equation of $A^n$, whose local solution is given by
Eq.\fr{eqn:linear_RGT}.
From Eq.\fr{eqn:linear_RGT}, we have
\beq
A^{n+1}-A^{n}=\Bigg(-i\ve\frac{J}{\sin\theta}
                     -\ve^2\frac{J^2}{2\sin^2\theta}
                 \bigg(2n+1+i\frac{\cos\theta}{\sin\theta}\bigg)
                 \Bigg)A+{\cal O}(\ve^3),
\lb{eqn:linear_nrg1}
\eeq
where $A$ should be
expressed in terms of $A^n$. This is done by taking
the inversion of the renormalization transformation \fr{eqn:linear_RGT}
iteratively
\beq
A=\bigg(1+i\ve\frac{Jn}{\sin\theta}+{\cal O}(\ve^2)\bigg)A^n.
\lb{eqn:linear_invA}
\eeq
Substituting \fr{eqn:linear_invA} into \fr{eqn:linear_nrg1}, we obtain
the following RG equation (RG map) up to ${\cal O}(\ve^2)$
\beq
A^{n+1}=\Bigg(1+\frac{-i\ve J}{\sin\theta}
          +\frac{1}{2!}\bigg(\frac{-i\ve J}{\sin\theta}\bigg)^2
          -i\ve^2\frac{J^2\cos\theta}{2\sin^3\theta}\Bigg)A^n
+{\cal O}(\ve^3),
\lb{eqn:linear_nRG}
\eeq
of which solution is
\beq
A^n=\Bigg(1+\frac{-i\ve J}{\sin\theta}
          +\frac{1}{2!}\bigg(\frac{-i\ve J}{\sin\theta}\bigg)^2
          -i\ve^2\frac{J^2\cos\theta}{2\sin^3\theta}
          +{\cal O}(\ve^3)\Bigg)^n A^0
.\lb{eqan:linear_nrg_sol}
\eeq

On the other hand, from Eq. \fr{eqn:linear_exact_x_sol},
we have $A^n$ exactly as
\beq
A^n=A^0\exp\Bigg[i\Bigg(\ve\frac{-J }{\sin\theta}
                -\ve^2\frac{\cos\theta}{2\sin\theta}
                          \bigg(\frac{J}{\sin\theta}\bigg)^2
                          +\cdots \Bigg)n\Bigg].
\lb{eqn:linear_exact_sol}
\eeq
Notice that $|A^n|^2$ is an exact constant of motion while
it is merely an approximate conserved quantity of
the (trancated) RG map \fr{eqn:linear_nRG}.
The symplectic structure is also not exactly preserved in the
RG map, that is, for the truncated RG map \fr{eqn:linear_nRG}
up to ${\cal O}(\ve^k)$, we have
$$
dA^{n+1}\land dA^{*~n+1}-dA^n\land dA^{*~n}={\cal O}(\ve^{(k+1)})
\ne 0,
$$
where $k=1,2,\cdots$. $A^{*}$ is complex conjugate to $A^n$ and
should also be a canonical conjugate to $A^n$ \cite{flow}.
Although this fault of the RG map vanishes in the limit of $\ve\to 0$.

In order to remedy a fault like this, we take advantage of a crucial
observation that Hamiltonian flows satisfy the following relation:
\beq
Z(t+\mu)=\bigg(1+\mu{\cal L}_H+\frac{\mu^2}{2!}{\cal L}^2_H+\bigg)Z(t)
=\exp(\mu{\cal L}_H)Z(t),
\lb{eqn:Liouville}
\eeq
where $t$ is the time variable, $\mu$ is a real number, $H$ is a Hamiltonian,
$Z$ is a canonical variable $(q_1,q_2,...,q_N,p_1,p_2,...,p_N)$ and
\beqa
{\cal L}_H Z&\equiv&\{Z,H\} \equiv
\sum_{j=1}^{N}\bigg(
\frac{\partial Z}{\partial q_j}\frac{\partial H}{\partial p_j}-
\frac{\partial Z}{\partial p_j}\frac{\partial H}{\partial q_j}
\bigg),
\lb{def:Liouville_op}\\
{\cal L}_H^2 Z&=&{\cal L}_H\bigg({\cal L}_H Z\bigg)=\{\{Z,H\},H\},\non
\eeqa
where ${\cal L}_H$ is the Liouville operator defined by the Poisson bracket.  
We can identify the relation \fr{eqn:Liouville} as the map by defining 
$Z^{n+1}\equiv Z(t+\mu)$, and  $Z^n\equiv Z(t)$:
\beq
Z^{n+1}=\Psi(Z^n;\mu),\qquad \Psi(Z^n;\mu)\equiv
\left.\exp(\mu{\cal L}_H)Z(t)\right|_{Z(t)\equiv Z^{n}}.
\lb{def:liouville_map}
\eeq
This is the symplectic mapping associted with the Hamiltonian system.
Identifying $\mu=\ve$ in \fr{def:liouville_map}, 
we can obtain both the Hamiltonian system and the associated symplectic 
mapping in the following steps:
\begin{itemize}
\item[1]
Exapanding a Hamiltonian $H$ in \fr{eqn:Liouville} in powers of $\ve$,  
$H=H^{(1)}+\ve H^{(2)}+\cdots$,
we obtain the following relation
\beqa
Z(t+\ve)&=&
\bigg(
1+\ve{\cal L}_H+\frac{\ve^2}{2!}{\cal L}^2_H+{\cal O}(\ve^3)\bigg)Z(t)\non\\
&=&\bigg\{
1+\ve{\cal L}_{H^{(1)}}+\ve^2
\bigg(
\frac{{\cal L}^2_{H^{(1)}}}{2!}+{\cal L}_{H^{(2)}}
\bigg)
+{\cal O}(\ve^3)
\bigg\}Z(t).
\lb{eqn:liouville_decomp}
\eeqa
\item[2]
$H^{(1)}$ can firstly be found by taking the limit 
$(A^{n+1}-A^n)/\ve\to dA/dt$, ($\ve\to 0$) in the naive RG map
by comparing the naive RG map with \fr{eqn:liouville_decomp}.  
Using $H^{(1)}$ and eqauting the naive RG map with the Liouville 
operator relation \fr{eqn:liouville_decomp}, we obtain the appropriate 
Hamiltonian $H=H^{(1)}+\ve H^{(2)}$.  
Similarly, we can obtain the higher Hamiltonian $H^{(3)},H^{(4)},...,$ order by order in $\ve$.
This procedure yields the approximate Hamiltonian. 
\item[3]
To obtain the symplecticity-preserving RG mapping, we discritize the 
continuous-time system obtained in the $2$nd step. 
The time-step should be chosen to be $\ve$.
\end{itemize}
 
In the present case, we take the limit $\ve\to 0$ in \fr{eqn:linear_nRG}, which gives
$$
\frac{dA}{dt}=-\frac{iJ}{\sin\theta}A=\frac{\partial H}{\partial A^*}
={\cal L}_{H^{(1)}}A
,\quad
\frac{dA^*}{dt}=-\frac{\partial H}{\partial A^*}={\cal L}_{H^{(1)}} A^*.
$$
Here $\frac{dA}{dt}$ is from $(A^{n+1}-A^n)/\ve$, and so on.
 Then we have 
$$
H^{(1)}=-i\frac{J|A|^2}{\sin\theta},\quad \mbox{and}\quad 
{\cal L}_{H^{(1)}}^2 A=\{\{A,H^{(1)}\},H^{(1)}\}=\frac{-J^2A}{\sin^2\theta}.
$$
At this stage, the Liouville operator relation is 
\beqa
&&\bigg\{
1+\ve{\cal L}_{H^{(1)}}
+\ve^2\bigg(\frac{{\cal L}_{H^{(1)}}^2}{2}+{\cal L}_{H^{(2)}}
\bigg)
\bigg\}A\non\\
&&=
A+\ve\frac{-iJA}{\sin\theta}+\ve^2
\bigg(
\frac{-J^2A}{2\sin^2\theta}+\frac{\partial H^{(2)}}{\partial A^*}
\bigg).\non
\eeqa
Equating this to the right hand side of the naive RG map \fr{eqn:linear_nRG},
we have
%$H^{(2)}$ can be obtained by the 
$$
%H^{(1)}=\frac{-iJ|\tA|^2}{2\sin\theta},\qquad
H^{(2)}=\frac{-iJ^2 \cos\theta |A|^2}{2\sin^3\theta}.
$$    
The Hamiltonian $H=H^{(1)}+H^{(2)}$ leads to the following equation:
$$
\frac{dA}{dt}=\frac{-iJ}{\sin\theta}A+\ve\frac{-iJ^2\cos\theta}{2\sin^3\theta}
A=\frac{\partial H}{\partial A},\qquad  \frac{dA}{dt}
=-\frac{\partial H}{\partial A^*}.
$$
The solution of the continuous-time system is
$$
A(t)=A(0)\exp
\bigg\{
i
\bigg(
\frac{-J}{\sin\theta}+\ve\frac{-J^2\cos\theta}{2\sin\theta^3}
\bigg)t
\bigg\},
$$
which yields the symplecticity-preserving RG map:
$$
A^{n+1}=A^n\exp
\bigg\{
i\ve
\bigg(
\frac{-J}{\sin\theta}+\ve\frac{-J^2\cos\theta}{2\sin^3\theta}
\bigg)
\bigg\}.
$$  
Here the time-step is chosen to be $\ve$.
Then we recover the exact solution of 
\fr{eqn:linear}, is \fr{eqn:linear_exact_sol}, in terms of the original variable $x^n$.

Finally, the present method can be summarized in the following diagram:\\

\noindent
\underline{A symplecticity-preserving RG method for symplectic maps}

\vspace*{0.9cm}
\hspace*{1.2cm}{ [discrete-time ]}\qquad 
\hspace*{3.7cm}
 {[continuous-time]}
\beqa
\begin{array}{ccc}
&&\\
\mbox{Symp. maps}~~&%\mapright{\mbox{$\tau\to 0$}}
&\\%\mbox{Cano. Eqs.}\\
{ \lmapdown{\mbox{naive RG}~}}&&\\%~{\rmapdown{ naive RG}}\\
\mbox{ naive RG maps}
&&\\
(\mbox{dissipative maps})&
\mapright{
\mbox{Liouville operator}}
&\mbox{RG Eqs.}\\
&&(\mbox{ canonical Eqs. })\\
\mbox{Symp.-Pres. RG maps}~~
&\mapleft{\mbox{Symp.-Pres. discretizations}}& \non
\end{array}
\eeqa

%%%%%%%%%%%%%%%%%%%%%%%%%%%%%%%%%%%%%%%%%%%%%%%%%%%%
\section{Two-dimensional Nonlinear Symplectic Map}
%%%%%%%%%%%%%%%%%%%%%%%%%%%%%%%%%%%%%%%%%%%%%%%%%%% 
\subsection{Non-resonant case}
%%%%%%%%%%%%%%%%%%%%%%%%%%%%%%%%%%%%%%%%%%%%%55
\qquad
Let us analyze the weakly nonlinear symplectic map
$(x^n,y^n)\mapsto (x^{n+1},y^{n+1})$:
\beqa
x^{n+1}&=&x^n+y^{n+1}\non\\
y^{n+1}&=&y^n-ax^n+2\ve J(x^{n})^3,\non
\eeqa
or
\beq
L_\theta x^n=\ve 2 J (x^n)^3,
\lb{eqn:nlin}
\eeq
where $\ve$ is the small parameter, $J$ is the real parameter, 
$\theta$ and $L_\theta x^n$ are 
defined in \fr{def:theta} and \fr{def:L} respectively.
Expanding $x^n$ as a power series of $\ve$  
$$
x^n=x^{(0)n}+\ve x^{(1)n}+x^{(2)n}+{\cal O}(\ve^3),
$$
we have
\beqa
L_\theta x^{(0)n}=0,\quad
L_\theta x^{(1)n}=2J(x^{(0)n})^3,\quad
L_\theta x^{(2)n}=6J(x^{(0)n})^2x^{(1)n}.
\lb{eqn:nlin_neqn}
\eeqa
and the solution to the perturbed equations to ${\cal O}(\ve^2)$
are given by 
\beqa
x^{(0)n}&=&A\e^{i\theta n}+\cc
\lb{eqn:nlin_n0}\\
x^{(1)n}&=&\frac{-3i|A|^2AJ}{\sin\theta}n\e^{i\theta n}+\frac{JA^3}{\cos 3\theta-\cos\theta}\e^{3i\theta n}+\cc
\lb{eqn:nlin_n1}\\
x^{(2)n}&=&
\bigg\{
\frac{-9}{2}\frac{J^2|A|^4A}{\sin^2\theta}n^2-i\frac{J^2|A|^4A}{\sin\theta}
\bigg(
\frac{3}{\cos 3\theta-\cos\theta}+\frac{9\cos\theta}{2\sin^2\theta}
\bigg)n
\bigg\}\e^{i\theta n}\non\\
&&+\bigg\{
\frac{-9iJ^2|A|^2A^3}{(\cos 3\theta-\cos\theta)\sin\theta}n+
\frac{J^2|A|^2A^3}{2(\cos 3\theta-\cos\theta)^2}
\bigg\{
12-18\frac{\sin 3\theta}{\sin\theta}
\bigg\}
\bigg\}\e^{3i\theta n}\non\\
&&+\bigg\{
\frac{3JA^5}{(\cos 5\theta-\cos\theta)(\cos 3\theta-\cos\theta)}
\bigg\}\e^{5i\theta n}+\cc.
\lb{eqn:nlin_n2}
\eeqa
Here $A(\in\bC)$ is the integration constant. 
To avoid resonant secular terms, we assume in addition that
$$
\cos\theta\neq\cos3\theta,\qquad \cos\theta\neq\cos5\theta. 
$$
The construction of the reduced map in a case of near resonance 
is considered in subsection 3.2. 
In order to remove the secular 
terms in the coefficient of the fundermental harmonic $(\exp(i\theta n))$, 
we introduce the renormalization transformation $A\mapsto A^n$:
\beqa
A^n&\equiv& A+\ve\frac{-3i|A|^2AJ}{\sin\theta}n\non\\
&&+\ve^2
\bigg\{
\frac{-9}{2}\frac{J^2|A|^4A}{\sin^2\theta}n^2\non\\
&&\qquad -i\frac{J^2|A|^4A}{\sin\theta}
\bigg(
\frac{3}{\cos 3\theta-\cos\theta}+\frac{9\cos\theta}{2\sin^2\theta}
\bigg)n
\bigg\}.
\lb{eqn:nlin_RGT}
\eeqa
Following the same procedure as that in the preceding section, we derive the
naive RG map from \fr{eqn:nlin_RGT}:
\beqa
A^{n+1}&=&A^n+\ve\frac{-3iJ}{\sin\theta}|A^n|^2A^n+\ve^2
\bigg\{\frac{1}{2!}
\bigg(\frac{-3iJ}{\sin\theta}|A^n|^2
\bigg)^2A^n\non\\
&&-\bigg(\frac{9i\cos\theta}{2\sin^3\theta}J^2
+\frac{3iJ^2}{\sin\theta(\cos3\theta-\cos\theta)}
\bigg)|A^n|^4 A^n
\bigg\},
\lb{eqn:nlin_nRG}
\eeqa
which breaks the symplectic symmetry. This RG map should be recovered 
the symplectic symmetry. 
%To obtain the symplecticity-preseriving RG map
%To recover the symplectic symmetry, we equate the naive RG map to the 
%Liouville operator relation \fr{def:Liouville_op}. 
The naive RG map can be written in terms of the real variables 
$A^n=A_1^n + i A_2^n$, 
\beqa
A_1^{n+1}&=&A_1^n+\ve\frac{3J}{\sin\theta}(A_1^{n2}+A_2^{n2})A_2^n\non\\
&&+\ve^2\bigg[
\frac{-1}{2!}\bigg\{\frac{3J}{\sin\theta}(A_1^{n2}+A_2^{n2})\bigg\}^2 A_1^n
\non\\
&&+\bigg\{ \frac{9\cos\theta}{2\sin^3\theta}
+\frac{3}{\sin\theta(\cos3\theta-\cos\theta)}\bigg\}J^2(A_1^{n2}+A_2^{n2})^2 A_2^n
\bigg],
\lb{eqn:nlin_nRG_A1}\\
A_2^{n+1}&=&A_2^n+\ve\frac{-3J}{\sin\theta}(A_1^{n2}+A_2^{n2})A_1^n\non\\
&&+\ve^2\bigg[
\frac{-1}{2!}\bigg\{\frac{3J}{\sin\theta}(A_1^{n2}+A_2^{n2})\bigg\}^2 A_2^n
\non\\
&&-\bigg\{ \frac{9\cos\theta}{2\sin^3\theta}
+\frac{3}{\sin\theta(\cos3\theta-\cos\theta)}\bigg\}J^2(A_1^{n2}+A_2^{n2})^2 A_1^n
\bigg].
\lb{eqn:nlin_nRG_A2}
\eeqa
The $(\ve\to 0)$-limit yields 
\beqa
\frac{dA_1}{dt}&=&\frac{3J}{\sin\theta}\bigg(A_1^2+A_2^2\bigg)A_2
=\frac{\partial H}{\partial A_2}={\cal L}_{H^{(1)}}A_1,\non\\
\frac{dA_2}{dt}&=&-\frac{3J}{\sin\theta}\bigg(A_1^2+A_2^2\bigg)A_1
=-\frac{\partial H}{\partial A_1}={\cal L}_{H^{(1)}}A_2,\non
\eeqa
where 
$$
H^{(1)}=\frac{3J(A_1^2+A_2^2)^2}{4\sin\theta}.
$$
This Hamiltonian generates the relations:
\beqa
&&
\bigg\{
1+\ve{\cal L}_{H^{(1)}}+\ve^2
\bigg(
\frac{{\cal L}^2_{H^{(1)}}}{2!}+{\cal L}_{H^{(2)}}
\bigg)
\bigg\}A_1(t)\non\\
&&=
A_1+\ve\frac{3J}{\sin\theta}(A_1^2+A_2^2)A_2\non\\
&&\qquad +\ve^2
\bigg\{
\frac{-1}{2!}\bigg(\frac{3J}{\sin\theta}\bigg)^2\bigg(A_1^2+A_2^2\bigg)^2A_1
+\frac{\partial H^{(2)}}{\partial A_2}
\bigg\},\lb{eqn:nlin_A1_second}\\
&&
\bigg\{
1+\ve{\cal L}_{H^{(1)}}+\ve^2
\bigg(
\frac{{\cal L}^2_{H^{(1)}}}{2!}+{\cal L}_{H^{(2)}}
\bigg)
\bigg\}A_2(t)\non\\
&&=
A_2+\ve\frac{-3J}{\sin\theta}(A_1^2+A_2^2)A_1\non\\
&&\qquad +\ve^2
\bigg\{
\frac{-1}{2!}\bigg(\frac{3J}{\sin\theta}\bigg)^2\bigg(A_1^2+A_2^2\bigg)^2A_2
-\frac{\partial H^{(2)}}{\partial A_1}
\bigg\}.\lb{eqn:nlin_A2_second}
\eeqa
According to the general procedure, 
$H^{(2)}$ can be obtained by equating 
them to \fr{eqn:nlin_nRG_A1}--\fr{eqn:nlin_nRG_A2}. 
That is, 
$$
H^{(2)}=\bigg\{
\frac{9\cos\theta}{2\sin\theta^3}+\frac{3}{\sin\theta(\cos3\theta-\cos\theta)}
\bigg\}
\frac{J^2(A_1^2+A_2^2)^3}{6}.
$$
Then we obtain the 
approximate Hamiltonian $H=H^{(1)}+\ve H^{(2)}$
\beqa
H&=&\alpha (A_1^2+A_2^2)^2+\beta (A_1^2+A_2^2)^3.\non\\
\alpha&\equiv&\frac{3J}{4\sin\theta},\qquad 
\beta\equiv\ve\bigg\{
\frac{9\cos\theta}{2\sin^3\theta}+\frac{3}{\sin\theta(\cos3\theta-\cos\theta)}
\bigg\}\frac{J^2}{6}.\non
\eeqa
The canonical transformation  $dA_1\wedge dA_2=d\Theta\wedge dI$:
$$
A_1={\sqrt 2I}\sin\Theta,\qquad A_2={\sqrt 2I}\cos\Theta,
$$
gives us the simple canonical equations:
\beqa
\frac{d\Theta}{dt}&=&\frac{\partial H}{\partial I}=8\alpha I+24\beta I^2,\non\\
\frac{dI}{dt}&=&-\frac{\partial H}{\partial \Theta}=0.\non
\eeqa
In addition to this expression, there exists the exponential form
\beqa
A&=&A_1+iA_2\non\\
 &=&\sqrt{2I(0)}(\sin\Theta+i\cos\Theta)\non\\
 &=&\sqrt{2I(0)}i(\cos\Theta-i\sin\Theta)\non\\
 &=&\sqrt{2I(0)}\exp\bigg(-i\bigg(8\alpha I(0)+24\beta I(0)^2\bigg)t
-i\theta (0)+i\pi/2\bigg).\non
\eeqa
Therefore, the solution of ($\ve\to 0$)-system which is written in the form of
exponential function is 
\beqa
A(t)&=&A(0)\exp
\bigg\{-it\bigg(
8 \frac{3J}{4\sin\theta}\frac{|A(0)|^2}{2}\non\\
&&+24 \ve\bigg\{ \frac{9\cos\theta}{2\sin^3\theta}
+\frac{3}{\sin\theta(\cos3\theta-\cos\theta)}\bigg\}\frac{J^2}{6}\frac{|A(0)|^4}{4}
\bigg)\bigg\}\non\\
&=&A(0)\exp
\bigg\{-it\bigg(
 \frac{3J}{\sin\theta}|A(0)|^2\non\\
&&+\ve\bigg\{ \frac{9\cos\theta}{2\sin^3\theta}
+\frac{3}{\sin\theta(\cos3\theta-\cos\theta)}\bigg\}J^2|A(0)|^4
\bigg)\bigg\},\non
\eeqa
where we use the relation $\sqrt{ 2I(t)}=|A(t)|=$ const.
To construct the reduced map, we discretize the Hamiltonian flow.
Identifying 
$$
A^{n+1}\equiv A(t+\ve),\qquad A^n\equiv A(t),
$$
yields a symplecticity-preserving RG map
\beqa
A^{n+1}&=&A^n\exp
\bigg[
i\ve
\bigg\{
\frac{-3J|A^n|^2}{\sin\theta}\non\\
&&+\ve J^2|A^n|^4
\bigg(
-\frac{9\cos\theta}{2\sin^3\theta}-\frac{3}{\sin\theta(\cos 3\theta-\cos\theta)}
\bigg)
\bigg\}
\bigg].
\lb{eqn:nlin_exponentiatedRG}
\eeqa
This map can also be 
obtained by the exponentiation RG method \cite{GN01JPSJ}.
%%%%%%%%%%%%%%%%%%%%%%%%%%%%%%%%%%%%%%%%%%%%%%%%%%% 
\subsection{Resonant case}
%%%%%%%%%%%%%%%%%%%%%%%%%%%%%%%%%%%%%%%%%%%
Let us consider a case of resonance in the mapping \fr{eqn:nlin}.
The solution to the perturbation equation \fr{eqn:nlin_neqn}  is 
obtained in the form \fr{eqn:nlin_n0}--\fr{eqn:nlin_n2} assuming that the 
parameter $\theta$ is far from the resonance with $\cos\theta=\cos3\theta$.
%The paper \cite{TD03} has reported that exponentiated RG maps cannot give 
%correct approximations near resonances.
Although resonant islands are important structures in symplectic maps,
they have never been analyzed by the RG method.
Here we demonstrate how our symplecticity-preserving RG method works 
near a resonant island. A similar procedure can be 
performed near other ones.

Let us expand $\theta$ near the resonance
$$
\theta=\frac{\pi}{2}+\ve\theta^{(1)}+\ve^2\theta^{(2)}+\cdots.
$$
Eq. \fr{eqn:nlin} can then be expressed in the form
$$
L_{\pi/2}x^n=\ve\bigg(2J(x^n)^3-2\theta^{(1)}x^n\bigg)-2\ve^2\theta^{(2)}x^n,
$$
where $L_{\pi/2} x^n$ is defined by 
$$
L_{\pi/2}x^n\equiv x^{n+1}+x^{n-1}.
$$   
The perturbation expansion yields 
\beqa
L_{\pi/2}x^{(0)n}&=&0,\non\\
L_{\pi/2}x^{(1)n}&=&2J(x^{(0)n})^3-2\theta^{(1)}x^{(0)n},\non\\
L_{\pi/2}x^{(2)n}&=&6J(x^{(0)n})^2x^{(1)n}-2\theta^{(1)}x^{(1)n}-2\theta^{(2)}
x^{(0)n}.\non
\eeqa
These Eqns can be solved, yielding the result
\beqa
x^{(0)n}&=&Ai^n+\cc\non\\
x^{(1)n}&=&(-i)i^n
n\bigg[
J(A^{*3}+3|A|^2A)-\theta^{(1)}A
\bigg]+\cc\non\\
x^{(2)n}&=&i^n n^2
\bigg[
\frac{3}{2}J^2\bigg(-2|A|^4A+|A|^2A^{*3}+A^5\bigg)\non\\
&&+J\theta^{(1)}
\bigg(3|A|^2A-A^{*3}\bigg)-
\frac{\theta^{(1)2}}{2}A
\bigg]+i^nni\theta^{(2)}A+\cc.\non
\eeqa
Here $A(\in\bC)$ is the integration constant.

As before, we define the renormalization transformation by
\beqa
A^n&\equiv& A+\ve (-i)n\bigg\{J(A^{*3}+3|A|^2A)-\theta^{(1)} A\bigg\}\non\\
&&+\ve^2
n^2
\bigg\{
\frac{3}{2}J^2(-2|A|^4A+|A|^2A^{*3}+A^5)
+J\theta^{(1)}
(3|A|^2A-A^{*3})-\non\\
&&\qquad \frac{\theta^{(1)2}}{2}A
\bigg\}+\ve^2 ni\theta^{(2)}A+\cc.\non
\eeqa
Taking into account the expression
$$
A=
A^n+\ve in\bigg\{J\bigg( (A^{*~n})^3+3|A^n|^2A^n\bigg)-\theta^{(1)} A^n\bigg\},
$$
which relates the amplitude $A$ to the renormalization variable $A^n$, 
we obtain the naive RG map 
\beqa
A_1^{n+1}&=&A_1^n+\ve\bigg(4J(A_2^n)^3-\theta^{(1)}A_2^n\bigg)\non\\
&&+\ve^2
\bigg\{
-24J^2(A_1^n)^3(A_2^n)^2+2J\theta^{(1)}\bigg( (A_1^n)^3+3A_1^n (A_2^n)^2\bigg)
\non\\
&&\qquad-\frac{\theta^{(1)2}}{2}A_1^n-\theta^{(2)}A_2^n\bigg\},
\lb{eqn:nlin_res_nRGA1}\\
A_2^{n+1}&=&A_2^n+\ve\bigg(-4J(A_2^n)^3+\theta^{(1)}A_1^n\bigg)\non\\
&&+\ve^2\bigg\{
-24J^2 (A_1^n)^2  (A_2^n)^3+2J\theta^{(1)}
\bigg( (A_1^n)^3+3 (A_1^n)^2 A_2^n\bigg)
\non\\
&&\qquad -\frac{\theta^{(1)2}}{2}A_2^n-\theta^{(2)}A_1^n
\bigg\}.
\lb{eqn:nlin_res_nRGA2}
\eeqa
Here $A^n=A_1^n+iA_2^n$, and $A_1^n,A_2^n$ are real variables.
$H^{(1)}$ can be determined by taking the limit $\ve\to 0$ 
\beqa
\frac{dA_1}{dt}&=&4JA_2^3-\theta^{(1)}A_2=\frac{\partial H^{(1)}}{\partial A_2},\non\\
\frac{dA_2}{dt}&=&-4JA_1^3+\theta^{(1)}A_1=-\frac{\partial H^{(1)}}{\partial A_1},\non\\
H^{(1)}(A_1,A_2)&=&(JA_1^4-\theta^{(1)}A_1^2/2)+(JA_2^4-\theta^{(1)}A_2^2/2).\non
\eeqa  
The smplecticity-preserving RG procedure gives the expression for 
$H^{(2)}$. Identifying $A^{n+1}$ in 
\fr{eqn:nlin_res_nRGA1}--\fr{eqn:nlin_res_nRGA2} as 
\beqa
&&A(t)+\ve{\cal L}_H A(t)+\frac{\ve^2{\cal L}^2_H}{2!}A(t)\non\\
&&=
A(t)+\ve\{ A(t),H^{(1)}\}+\ve^2
\bigg(
\{A(t),H^{(2)}\}+\frac{1}{2!}\{\{ A(t),H^{(1)}\},H^{(1)}\}
\bigg),\non
\eeqa
then the trancated Liouville operator relations are 
\beqa
A_1(t+\ve)&=&A_1(t)+\ve\bigg(4JA_2^3-\theta^{(1)}A_2\bigg)\non\\
&&+\ve^2
\bigg[
-24J^2A_1^3 A_2^2 +2J\theta^{(1)}\bigg(A_1^3+3A_1 A_2^2\bigg)\non\\
&&\qquad -\frac{\theta^{(1)2}}{2}A_1-\theta^{(2)}A_2+\frac{\partial H^{(2)}}{\partial A_2}\bigg],
\lb{eqn:nlin_res_LiouA1}\\
A_2(t+\ve)&=&A_2 +\ve[-4JA_2^3+\theta^{(1)}A_1]\non\\
&&+\ve^2\bigg[
-24J^2A_1^{2}A_2^3+2J\theta^{(1)}\bigg(A_1^3 +3A_1^2A_2\bigg)\non\\
&&\qquad -\frac{\theta^{(1)2}}{2}A_2-\theta^{(2)}A_1
-\frac{\partial H^{(2)}}{\partial A_1}\bigg]
\lb{eqn:nlin_res_LiouA2}.
\eeqa
Compring \fr{eqn:nlin_res_LiouA1}--\fr{eqn:nlin_res_LiouA2} with 
\fr{eqn:nlin_res_nRGA1}--\fr{eqn:nlin_res_nRGA2}, we have the Hamiltonian
whose trajectory can approximatly interpolate the trajectory of 
the naive RG map,
$$
H=H^{(1)}+\ve H^{(2)}=J(A_1^4+A_2^4)-\frac{\theta^{(1)}+\ve\theta^{(2)}}{2}(A_1^2+A_2^2).
$$
This Hamiltonian system is integrable because of $1$ degree of freedom. 
However the solution may be complicated to write the analitic expression,
which yields the difficulty of finding a symplecticity-preserving RG map.
Instead of using the analytical expression to the Hamiltonian flow, we use 
a symplectic integrator, which is known to be a discretization method designed 
for preserving the symplecticity for Hamiltonian flows\cite{Yos93}.   

We split the Hamiltonian as $H_1+H_2$, not $H^{(1)}+\ve H^{(2)}$,
so that we use a symplectic integrator,
$$
H=H_1+H_2=\bigg(JA_1^4-\frac{\theta^{'(1)}}{2}A_1^2\bigg)
+\bigg(JA_2^4-\frac{\theta^{'(1)}}{2}A_2^2\bigg),\quad \theta^{'(1)}
\equiv\theta^{(1)}+\ve\theta^{(2)}.
$$
The Hamiltoinian $H_1$ provides the mapping
$$
\e^{\tau D_{H_1}}:\quad A_1(t+\tau)=A_1(t),\quad  
A_2(t+\tau)=A_2(t)+\bigg(-4JA_1^3(t)+\theta^{'(1)}A_1(t)\bigg)\tau.
$$
Here $\tau$ is a real number. Similarly, $H_2$ provides
$$
\e^{\tau D_{H_2}}:\quad A_1(t+\tau)=A_1(t)+\bigg(4JA_2^3(t)-\theta^{'(1)}A_2(t)\bigg)\tau,\quad  
A_2(t+\tau)=A_2(t).
$$  
Therefore we can obtain the reduced symplectic map 
\beqa
A_1^{n+1}&=&A_1^n
+4J\ve\bigg[\bigg\{A_2^n+\frac{\ve}{2}(-4JA_1^{n3}+\theta^{'(1)}A_1^n)\bigg\}^3
\non\\
&&-\theta^{'(1)}\bigg\{A_2^n+\frac{\ve}{2}(-4JA_1^{n3}+\theta^{'(1)}A_1^n)\bigg\}
\bigg],
\lb{eqn:nlin_res_SP_A1}\\
A_2^{n+1}&=&A_2^{n}+\ve\bigg(-4JA_1^{n3}+\theta^{'(1)}A_1^n\bigg)\non\\
&&+\frac{\ve}{2}\bigg(-4JA_1^{n+1~3}+\theta^{'(1)}A_1^{n+1}\bigg).
\lb{eqn:nlin_res_SP_A2}
\eeqa
Here we have taken the following symplectic integrator:
$$
\e^{\ve D_H}=\exp(\ve \frac{D_{H_1}}{2})\exp(\ve D_{H_2})\exp(\ve \frac{D_{H_1}}{2})+{\cal O}(\ve^3).
$$

%%%%%%%%%%%%%
%\pagebreak 
%5%%%%%%%%%%%%

%%%%%%%%%%%%%%%%%%%%%%%%%%%%%%%%%%%%%%%%%%%%%%%%%%% 
\subsection{Numerical results in case of a resonance}
%%%%%%%%%%%%%%%%%%%%%%%%%%%%%%%%%%%%%%%%%%%%%%%%%%%
In this subsection we present illustrative numerical results for our 
symplecticity-preserving RG method in case of near the resonance with 
$\cos\theta=\cos3\theta$.
To show that %our Liouville operator appoach to symplecticity-preserving 
%RG method approximates original symplectic maps, 
our RG method can successfully be applied to study a resonant structure
for the 2-dimensional symplectic map \fr{eqn:nlin},   
we use both the reduced maps 
\fr{eqn:nlin_res_SP_A1}--\fr{eqn:nlin_res_SP_A2} and 
\fr{eqn:nlin_exponentiatedRG} up to ${\cal O}(\ve)$. 

%%%%%%%%%%%%%%%%%%%%%%%%%%%%%%%%%%%%%%%%%%%%%%%%%
\begin{figure}[h]
\begin{center}
\includegraphics[width=4.4cm]{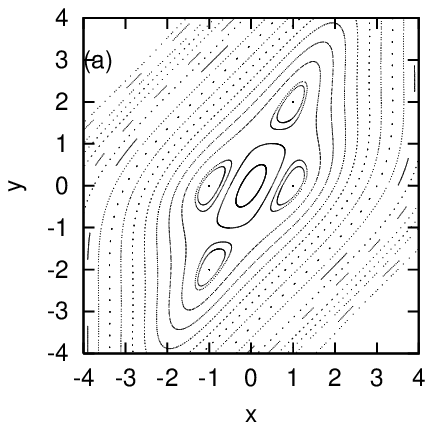}
\includegraphics[width=4.4cm]{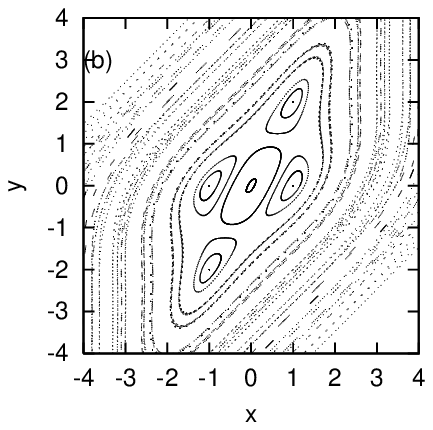}
\includegraphics[width=4.4cm]{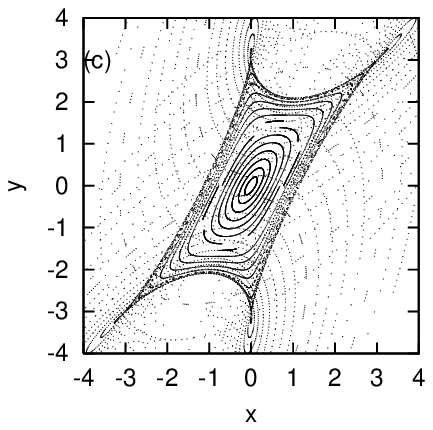}
\caption{Phase portraits of the 2-dimensional symplectic map model,
with the parameters are $\ve=0.01,J=1.0$, and $\theta^{(1)}=1.0$: (a) the original map [Eq. \fr{eqn:nlin}], 
(b) the Liouville operator approach to the RG method
[Eq. \fr{eqn:nlin_res_SP_A1}--\fr{eqn:nlin_res_SP_A2} up to ${\cal O}(\ve)$ ], 
(c) the exponentiated RG method [Eq.\fr{eqn:nlin_exponentiatedRG} up to 
${\cal O}(\ve)$].}
\end{center}
\end{figure}
%%%%%%%%%%%%%%%%%%%%%%%%%%%%%%%%%%%%%%%%%%%%%%%%%%

%\vspace*{-3.4cm} 
%\hspace*{1.5cm}(a),\hspace*{3cm} (b),\hspace*{3cm} (c) \\
%
%\vspace*{3.4cm} 

%%%%%%%%%%%%%%%%%%%%%%%%%%%%%%%%%%%%%%%%%%%%%%%%%
\begin{figure}[h]
\begin{center}
\includegraphics[width=4.4cm]{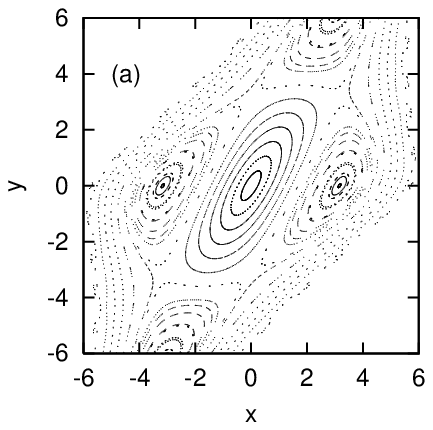}
\includegraphics[width=4.4cm]{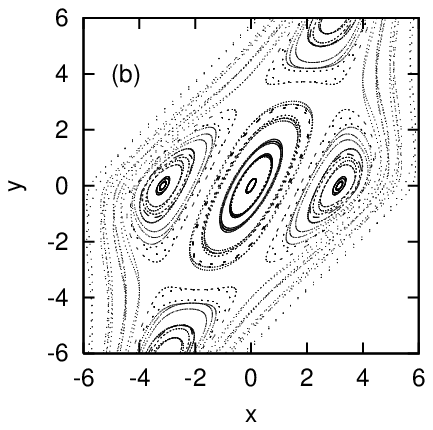}
\includegraphics[width=4.4cm]{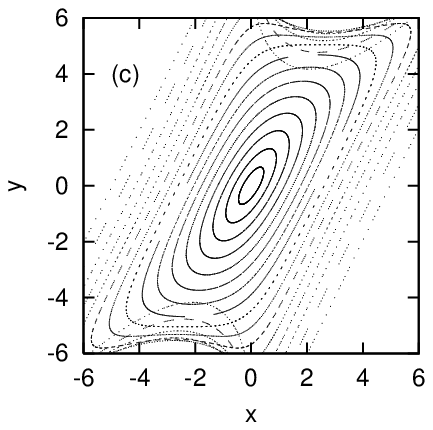}
\caption{Phase portraits of the 2-dimensional symplectic map model,
with the parameters are $\ve=0.01,J=1.0,$ and $\theta^{(1)}=10.0$: (a) the original map [Eq. \fr{eqn:nlin}], 
(b) the Liouville operator approach to the RG method
[Eq. \fr{eqn:nlin_res_SP_A1}--\fr{eqn:nlin_res_SP_A2} up to ${\cal O}(\ve)$ ], 
(c) the exponentiated RG method [Eq.\fr{eqn:nlin_exponentiatedRG} 
up to ${\cal O}(\ve)$].}
\end{center}
\end{figure}
%%%%%%%%%%%%%%%%%%%%%%%%%%%%%%%%%%%%%%%%%%%%%%%%%%

%\vspace*{-3.4cm} 
%\hspace*{1.5cm}(a),\hspace*{3cm} (b),\hspace*{3cm} (c) \\
%
%\vspace*{3.4cm} 

In figures 1--2 the phase portraits to the map near the resonance with
$\cos\theta=\cos3\theta$, are depicted. 
Although the exponentiated RG map 
agrees well with the exact numerical result for near the origin in the phase space, 
it derivates considerably from 
the exact result near the resonance point. In contrast, the map obtained by 
the Liouville operater approach to a symplecticity-preserving RG method can globally give a good
approximation even in 
the case of the $1$st RG approximation.     
Detailed studies of  such  resonant island structure by means of the 
present RG method will be
published elsewhere for other maps including the H\'enon map.

%%%%%%%%%%%%%%%%%%%%%%%%%%%%%%%%%%%%%%%%%%%%%%%%%%% 
\section{Conclusions}
%%%%%%%%%%%%%%%%%%%%%%%%%%%%%%%%%%%%%%%%%%%%%%%%%%%
We present the Liouville operator approach to the RG method 
to preserve symplectic structures in RG maps near elliptic fixed points of 
symplectic discrete systems. The symplecticity-preserving procedure
is accomplished by comparing naive RG maps with the Liouville operator relation
order by order in the small parameter and gives symplectic maps, which successfully describes the 
long-time asymptotic behavior of the original systems.  
Although the exponentiation procedure has not been able to be applied to 
study resonant islands, the Liouville operator approach to symplecticity 
preservation has given correctly reduced maps. 
%This means that we have successfully 
%obtained a reduced map even when a resonant island chain appears. 
Furthermore the advantages of  
this new method are  
not only a time-step parameter is not needed, 
but also reduced maps may be able to be explicit schemes.       

It is easy to see that the present symplecticity-preserving method is also 
applicable to general weakly-nonlinear symplectic maps.
Other symplecticity-preserving RG maps are to 
be studied in future.

%%%%%%%%%%%%%%%%%%%%%%%%%%
\section*{Acknowledgement}
%%%%%%%%%%%%%%%%%%%%%%%%%%
One of the authors (S.G.) has been supported by 
a JSPS Fellowship for Young Scientists. 
The another author (K.N.) has been, in part, supported by a Grant-in-Aid for 
Scientific Research
(C) 13640402 from the Japan Society for the Promotion of Science.
%%%%%%%%%%%%%%%%%%%%%%%%%%%%%%%%%%%%%%%%%%%%%%%%%%%%%%%%%%%%%%%%

%
%%%%%%%%%%%%%%%%%%%
%\section*{Appendix}
%%%%%%%%%%%%%%%%%%%

%%%%%%%%%%%%%%%%%%%%%%%%%%%

%%%%%%%%%%%%%%%%%%%%%%

%%%%%%%%%%%%%%
\end{document}